\documentclass[modern]{aastex631}
\usepackage{graphicx}
\usepackage{mathtools}
\usepackage{xspace}
\graphicspath{ {./Figures/} }
\usepackage[multiple]{footmisc}
\journalinfo{}
\usepackage{color}

\def\eg{{\it e.g., }}

\def\ie{{\it i.e., }}

\def\gsim{~\rlap{$>$}{\lower 1.0ex\hbox{$\sim$}}}
\def\lsim{~\rlap{$<$}{\lower 1.0ex\hbox{$\sim$}}}
\usepackage[margin=1in]{geometry}

\def\vaibify{\texttt{\small{vaibify}}\xspace}

\begin{document}

\title{Integrating AI Coding Agents Into Scientific Research}

\author[0000-0001-6487-5445]{Rory Barnes}
\affiliation{Department of Astronomy, University of Washington, 3910 15th Avenue NE, Seattle, WA 98195, USA}

\begin{abstract}
\noindent New software tools have emerged that enable AI agents to quickly develop code that analyzes data and generates figures, but their mistakes can also harm a researcher's computer, require extensive resources for validation, and cause the results to be received more skeptically. These concerns could be mitigated if scientific software is developed in a secure environment with a system for monitoring file changes and well-defined standards for handling AI agents. This paper first describes how coding agents change eleven aspects of generating, verifying, and reproducing numerical results. A six-level scale is then proposed for classifying AI-assisted numerical experiments by their provenance, reproducibility, openness, oversight, and falsifiability (PROOF). The scale begins with locally self-consistent, version-controlled, human-verified results and culminates in a publicly accessible and curated archive of the files, change history, prompt transcripts, independent agent regeneration of the analysis, and multiple cryptographically verifiable attestations of regeneration by independent researchers. These aspects of AI-assisted scientific software development have been synthesized into a new open source software package called \vaibify that optimizes the agentic development of computational experiments ascending the PROOF Ladder. Docker containers sequester the agents while a GUI enables researchers to prompt command line agents, monitor the current status of local and remote files, and drive a project to PROOF Level 3 (reproducible). The methods described in this paper could help researchers reduce the time required to verify agent-written software and efficiently communicate trust in AI-assisted science.
\end{abstract}

\section{Introduction\label{sec:intro}}

The rise of powerful agentic AI coding tools, such as Anthropic's \texttt{Claude Code} and OpenAI's \texttt{codex}, offers the possibility of quickly generating scientific results due to the speed at which they write functioning code. Moreover, these tools have been trained on a large body of text describing modern statistical techniques and the scientific literature in general, so they can often quickly ``understand'' a research problem. AI coding agents thus have the potential to dramatically improve research in the physical sciences. This shift could significantly change how modern science is practiced, which raises a host of practical, ethical, and philosophical concerns \citep[see \eg][]{Hogg26} that will not be addressed here. This article instead describes a possible approach to developing, verifying, and archiving scientific code and data generated completely by AI models, including a brief presentation of a new software package called \vaibify that can sequester and monitor AI agents while they write scientific code and analyze its output.

While the promise of large language models (LLMs) is great, they also demonstrate stochastic behavior \citep{Ouyang25}, resulting in non-zero probabilities that code will be wrong \citep{OBrien25}, or, worse, critical files will be erased without authorization. Furthermore, scientists may trust the AI too much and not fully validate the numerical experiment, or not trust it enough to submit accurate and compelling research (an arguably worse outcome). Nonetheless, the speed at which they write useful code demands a consideration of how LLMs can be leveraged for cutting edge science, how their dangerous effects can be minimized, and how to maximize human trust in the code and data files they generate. When employing AI coding agents, the time bottleneck appears to be shifting from writing code to verifying code. These themes are explored in this article, leaving alone the harder (and perpetual) problem of properly interpreting results, \ie evaluating if outcomes are accurate and/or publishable.

For now at least, the interpretation of cutting edge results remains a feature of science that cannot be fully outsourced to AI. Even efforts in which AI agents fully conceive, execute, and write scientific papers, such as \texttt{Denario} \citep{VillaescusaNavarro25} and the \texttt{AI Scientist} \citep{Lu26}, still rely on human evaluation of the results. The premise of this article is that AI agents are valuable enough today to integrate into most research projects, but that researchers still play a foundational role by verifying that the AI's output is correct. To begin, 11 features of scientific software development that change when employing AI coding agents are presented ($\S$\ref{sec:amenable}), a six-step scale is then proposed that qualitatively measures the trustworthiness of an AI-assisted numerical experiment ($\S$\ref{sec:proof}), and finally a new software tool that embraces these principles to empower scientists to safely direct coding agents to build, test, maintain, and archive scientific code and its output is outlined (App.~\ref{sec:vaibify}).

\section{Properties of AI-Assisted Scientific Software Development\label{sec:amenable}}

While leveraging AI coding agents in scientific analyses is a nascent endeavor, some suggestions of how to use them are starting to emerge \citep[\eg][]{Chen25}. Safety, verification, and reproducibility appear to be the most critical aspects of AI-assisted research. The first is grounded in human nature: Nobody wants to use a tool that could cause permanent harm. The latter two assume that scientists strive to publish accurate information that their colleagues can trust and build upon. Other important aspects to consider are the speed to insight, the ability to archive digital artifacts that can be examined by future scientists, and procedures that prevent LLMs from deviating from established architectural decisions in the software's design. 

Most of these goals already align with the modern scientific method and recommendations for open science \citep{OpenScience18}, and pre-AI era efforts to organize and curate digital research artifacts \citep[\eg][]{Peng11,Sandve13,Stodden16,Wilkinson16,Brinckman19,Luger21}. Many key constraints on design decisions remain intact as well, such as how quantifiable, tractable, iterable, parallelizable, and archivable the project is. The abilities of AI agents, however, demand a re-evaluation of the methodology and priorities. This section presents a partial --- and personally verified --- list of how AI coding agents can affect decisions in the scientific software development lifecycle.

\subsection{Containable} 

Harnessing AI agents to write code and manage directory structures can be dangerous because poorly constructed prompts or reckless agents can cause serious and permanent harm to a user's computing environment. Even if they cause no harm, coding agents can introduce security vulnerabilities \citep{Pearce25}. While such events appear to be rare phenomena with modern agents, these failure modes can be catastrophic.  Thus, AI agents should only be able to access the files that are necessary for the research tasks in a computing environment with absolutely no access to critical files or machines. Moreover, a contained agent only has access to the data and software packages the researcher approves, so a fully contained agent is less likely to search for spurious information and insert dubious code into software packages.

\subsection{Translatable} 

At its core, an AI agent maps natural language prompts into source code and shell scripts. Code is more predictable than natural language \citep{Hindle12}, which is a major reason why natural language prompts can be so effectively translated into code. The approximate translations, however, still admit error and thus the accuracy with which an LLM can perform this translation partially dictates the value of the resultant code. Moreover, many projects are tens or hundreds of thousands of lines of code long, which can challenge an agent's context window, but certain styles can be more efficient than others, such as naming variables with full, descriptive words instead of arbitrary characters. 

In an ideal project, a design document paired with the same LLM harness always reproduces the same source code, {\it and vice versa.} While the return trip seems like a distant goal, in general, a strict set of semantic rules for the source code can improve translatability and speed-up code development. AI-developed code should follow strict style standards that minimize the ``perplexity,'' a measurement of how unusual a specific sequence of tokens is compared to the training corpus of the LLM \citep{Ray16}. More perplexing code, which doesn't adhere to style invariants, will be more challenging for an LLM to parse and predict the next token. In other words, the translation is more prone to errors.

\subsection{Localizable} 

The limited context windows of modern AI coding agents are best leveraged when concepts and scopes are confined to small blocks of code \citep{Ding23,Liu24}, which allows agents to more easily understand the goals of a function. Moreover, AI-assisted code should avoid global variables or other dependencies that may be spread across files. These limitations may be unavoidable for some projects, so in those cases comments should clearly explain the complex dependencies and/or larger scope. 

\subsection{Contestable} 

The rise of multiple, high-quality coding agents has presented the possibility of asking them to debate ideas and review each other's work, \ie they can contest each other's findings. In practice, the value of pitting agents against each other is mixed, with some studies suggesting that a well-formed prompt to a single agent is as effective as a poorly worded prompt to multiple agents \citep{Wang24,Smit24}. Thus, prompting a discussion between agents can produce results that are unlikely to be worse than a prompt to a single agent. But given the higher cost of inviting debate between agents, this approach should probably only be used for the most difficult challenges.  

\subsection{Decomposable} 

Current AI agents have limited context memory (but that is changing quickly), so only projects that can be split into small enough pieces are viable. Additionally, the dependencies between the components must be identifiable to ensure that one step is not relying on unverified data. In most research projects, these steps typically consist of the generation of data and the creation of plots, with some steps depending on the output of others. Decomposing problems into steps is, of course, a classic treatment of complex problems (a ``pipeline'') \citep[\eg][]{DiTommaso17,Molder21}, so this step in isolation is not groundbreaking; only the relation to context window size is new.

\subsection{Verifiable} 

Scientists must validate AI-assisted results before publishing. While in the past, scientists could rely on their own coding prowess to trust their experiment worked, AI-written code requires new approaches. In the traditional methodology, results are also verified by examining plots and/or raw data, which still works in the AI era \citep{OBrien25} and so should not be replaced. Crucially, however, AI agents can sometimes unexpectedly modify previously verified data, which can have downstream consequences. For example, a script in Step 2 of a project creates a data file that is used in Step 8, but then an agent subsequently modifies the Step 2 script without alerting the researcher. The outcome is that the Step 8 result is now out-of-date and the researcher doesn't know it. This example highlights that the timings of output file generation must be monitored to ensure quality control. 

Furthermore, testing structures, such as unit tests, appear to be a critical requirement for AI-generated code to ensure that new modifications do not break previously verified functionality \citep{Liu23}. If possible, all tests for a specific step in a pipeline should be generated by deterministic processing of the results, \eg a Python script, instead of a prompt to an AI agent. Invoking deterministic methods is, in general, always preferable to prompts when working with AI agents. 

More broadly, verification may now be replacing code development as the bottleneck for completing a project. Thousands of lines of fully functioning code can be written in just minutes, but the validation of those lines can now require days or weeks because the human developer must be convinced that the code is both accurate and sustainable. Streamlining code verification may be the next frontier in software development.

\subsection{Observable} 

All code should generate enough --- but just enough --- output for the outcome and any errors to be unambiguously interpreted. Excessive logging is difficult for humans to parse and identify crucial messages. For AI-generated code, this issue is even more important because both humans and agents may not have a memory of where the offending functions lie in the codebase. Thus, software should exit gracefully with a helpful message that guides humans and agents to the solution.

\subsection{Reproducible} 

All scientific work should strive for full reproducibility, but this goal takes on more urgency with AI-assisted research. In general, AI agents can easily regenerate results from repositories, write qualitative descriptions, and assess reproducibility \citep{Hu25}. The fundamentally new issue is AI stochasticity;  quantitative output should minimize dependencies on a specific LLM's output. Data, scripts, and output should be tracked and available for other researchers to investigate and recover claimed results, independent of AI and the original researchers. Statements regarding how AI was used, such as the specific tasks and model used, should be reported to help future researchers reconstruct how AI was inserted into the scientific method. These sorts of records could, for example, be useful if biases are later discovered in certain LLMs that could adversely affect how they wrote code.

\subsection{Falsifiable} 

Unit tests are explicitly designed to confirm an executable's behavior; they do not prove that incorrect code would fail. Nonetheless, developers often assume that they do, resulting in a false sense of security in terms of accuracy and reliability. This limitation can be circumvented if the unit tests themselves are tested to confirm that they catch known failure modes. Since researchers hypothesize that unit tests are catching failures, these intentional breakages can be thought of as ``falsification tests'' of the unit test hypothesis. This approach has typically been called ``mutation testing'' \citep{DeMillo78,Papadakis19} and requires a developer to intentionally change the code, \eg flipping a $<$ to a $>$ (a ``mutation''), and then checking that one or more unit tests fail. Because the changes are deliberate, I prefer the term ``falsification test'' to ``mutation test,'' which implies a certain amount of unpredictability. This proactive form of testing can be especially important when a single AI agent writes the code, writes the unit tests, and then evaluates the testing suite's success by inserting additional deterministic checks on the development. 

Mutation testing can be tedious for humans to write, but numerous packages have been developed to perform this task, such as \texttt{mutmut}\footnote{Publicly available at https://github.com/boxed/mutmut.} and \texttt{cosmic-ray}\footnote{Publicly available at https://github.com/sixty-north/cosmic-ray.}. Fortunately, the speed at which AI reads and writes code can be harnessed in this case to identify where to apply falsification tests. Furthermore, a recent analysis suggests that LLMs are better at generating ``mutants'' that mimic real bugs compared to traditional, deterministic techniques \citep{Wang26}. Hence directing an agent to generate falsification tests appears to be a promising avenue for maintaining code integrity.

Crucially, however, falsification testing can be ambiguous in some cases since intentionally broken code can still produce the expected outcomes (``equivalent mutants''), but the larger the fraction of written code that can be falsified, the more likely it is to function as expected and the less likely future changes will break existing functionality. Further complicating falsification testing procedures is that parallel testing is challenging because the code modifications from one falsification test can interfere with another if performed on the same git branch. Git worktrees can mitigate this issue, but falsification testing is likely the bottleneck in comprehensive testing suites.  

\subsection{Enforceable} 

All code includes arbitrary design decisions that are made as the code is developed. In well-written code, the design is generally optimized for modularity, readability, and consistency \citep{Ford17}. Modern LLMs have relatively short context windows compared to the size of legacy codes, which may result in newly instantiated agents making unexpected design choices unless rules are documented and enforced. Thus, AI-friendly code should have clear specifications for the software architecture, source code style, and network connections, \eg if paths to files are represented as relative or absolute. In practice, repositories should include tests that ensure that the code does not deviate from these invariants. Such an approach ensures that new agents are not able to break existing design decisions, even if the resulting output is correct, \ie the unit and falsification tests pass. 

\subsection{Replayable} 

Finally, the ideal project facilitates the complete reconstruction of the code and results by future researchers. Such a project would publish all agent instruction files (\eg AGENTS.md)  as well as the transcript of the conversation with the agents. A replayable project also records the LLM model(s) used with each prompt, with open-weight models' transparency making them the preference over closed models. 

The full prompt transcript offers several benefits to future researchers. First, it enables them to search for any misconceptions in the original researchers' prompts that may have introduced incorrect methods in the software. Second, it can help scientists who are new to using AI agents learn which prompt styles are most effective at generating functional code. Finally, its publication increases transparency, which will make it more likely that other researchers will trust fully AI-written code.

A full record of the AI's involvement is the ultimate constraint on its role in the project as it enables repeated reconstructions of the process and the opportunity to determine if independent LLMs would generate the same results. More broadly, a timestamped prompt history can be combined with the git history to help determine agents' motivations for all decisions. Some development decisions, however, will likely be made ``by hand'' by the researcher, and thus exist outside the replayability. Thus, true replayability should also include a ``supervision log'' that records all file changes with timestamps, regardless of who or what modified the file. A commitment to replayability requires significant effort, but it may be the only way to fully assess AI's role in scientific research.

One practical issue with replayability is that most codes undergoing active development today were started before widespread adoption of AI agents, but now employ them. Thus, the first modifications of a codebase by an AI agent may be uncertain and/or the developers may not have had the tools to record the prompts and/or didn't expect that it would be valuable to do so. The best way to handle these cases is not obvious, but a reasonable approach could be for the developers to acknowledge the situation and record when they began using agents. Some coding agents can be considered authors of specific git commits, which, if enabled, can at least provide a lower limit of the time since agents began modifying the code. Many AI companies store chat histories, so in principle it is possible to recover the prompt transcripts if they are willing to share them. Nevertheless, these legacy codes represent an edge case that will likely be important in the near-term as these codes begin to be modified by AI agents.

\subsection{Synthesizing the Properties}

The previous subsections explored how scientific software development changes when it is written by AI coding agents, focusing on a complete approach from conception to publishing to archiving to reproduction. Furthermore, the assumption is that the research is developed on a local machine, but also stored remotely. Critically, the discussion does not include accuracy; its scope is limited to the development of software and the verification that results are reproducible. 

The above list only guides researchers on how to incorporate AI into their research. Without clear rules and procedures for development and reproducibility, a skeptical reader of a manuscript describing AI-assisted research would reasonably toss it out if the authors did not fully explain how AI was used. The next section describes how the role of AI can be assessed in research.

\section{The PROOF Ladder\label{sec:proof}}

To evaluate how well a data science project controlled for the stochastic behavior of LLMs, this section offers a scale that qualitatively describes ever-stricter standards of confidence that AI has not inadvertently modified the files produced by a data analysis. The levels are somewhat arbitrary, but they draw heavily on other archival methods, such as the Supply-chain Level for Software Artifacts (SLSA) framework, the Association for Computing Machinery's v1.1 badging scheme\footnote{https://www.acm.org/publications/badging-terms}, and artwork certification, \ie industries in which provenance and continuous confirmation of authenticity (``custody'') are of paramount importance. These methods are similar to current norms (or aspirations) of most scientific studies, suggesting this approach is of general applicability. The scale is {\it not} intended to confirm that results are accurate representations of reality; it only addresses how well scientists monitored the role of AI in the research.

The PROOF Ladder draws most heavily from the SLSA initiative, which is a community effort that has established a checklist of standards for monitoring the digital infrastructure underpinning international supply chains. SLSA helps vendors confirm they're using software that has not been modified by malicious actors. While not a perfect match to scientific research, the analogy is clear: we seek to ensure the software we use is free of corruption introduced by AI agents. SLSA created standards for source code\footnote{https://slsa.dev/spec/v1.2/source-requirements\#basics} that appear most applicable here and identified 4 levels of source code confidence: 1) Use of a version control system; 2) a changelog that details source code provenance; 3) continuous technical control of the repository, \ie enforcement of the particular rules of an individual package, and 4) independent code review by multiple trusted persons. This ladder allows companies to measure the trustworthiness of their software ecosystem. An analogous scale for scientific software is presented below.

Given the pace of LLM model and harness development, this proposed scale may quickly become outdated, but it could provide some guidelines on how trustworthy an AI-assisted scientific result is. At the time of this writing, very few published studies have integrated AI agents into their experiment, so the vast majority of previous research is independent of AI coding agents. Future research, however, seems likely to be significantly impacted by coding agents, so the concepts presented here could provide a foundation for future research to safely and effectively leverage LLMs for cutting edge science and enable scientists to objectively evaluate the role of AI.

\subsection{Level 1: Self-Consistent\label{sec:proof-1}} 

The lowest level of the scale is a project whose code and data are locally consistent and under version control. Outcomes are fully validated by a clear set of rules, such as unit tests that cover the vast majority of the code (ideally $\ge 95$\%). A repository reaching Level 1 has crossed an important threshold as it marks a project nearing completion: the data are analyzed; the figures are approaching camera-ready; and the researchers are invested enough to monitor the results. They believe they have controlled the agents sufficiently that the results could be worth sharing with the broader community. However, the project is invisible to the rest of the world, and is therefore, by definition, {\it not} a reproducible scientific result.

\subsection{Level 2: Published\label{sec:proof-2}} 

The public release of data along with a clear and complete declaration of how the AI was used and controlled marks the ascension to Level 2. Maturing scientific projects tend to connect to one or more remote resources for both keeping their effort safe, as well as preparing for some sort of publication that presents the data analysis. These remote resources are natural locations to release data products associated with a paper and include the statement on AI usage. Note that projects that seek to reach Level 4 and/or 5 must include metadata and specific tolerances for output parameters with their publication, see $\S\S$~\ref{sec:proof-4}--\ref{sec:proof-5}. While publication obviously represents a critical step, the release does {\it not} explain how other researchers can exactly reproduce the published results.

\subsection{Level 3: Reproducible\label{sec:proof-3}} 

The next step in the ladder adds the completion and announcement of the instructions on how to reproduce results, ideally at the bit level, including an attestation that the original authors have in fact performed the reproduction. A Level 3 product provides a mechanism to collect all dependencies in a clean, well-defined environment and regenerate the results exactly as in the original experiment. A Level 3 repository should include hashes, such as SHA-256, that safely allow others to compare every byte of their recreation to every byte of the original investigation. A guarantee of byte-reproducibility for an arbitrary project may not be possible due to multi-threading, CPU settings, etc., but a Level 3 project will ideally allow most independent researchers to exactly reproduce a result if they use the same hardware (and its settings) as the original author. A reproducible result does {\it not} have to contain all the raw observational data necessary to reproduce the results, nor does it include a public record of how all the data and analysis software were assisted by AI.

\subsection{Level 4: Traceable\label{sec:proof-4}} 

The next level adds the complete recording of the sequence of events from the first commit to the curation of a long-term storage repository, including the raw data, analysis scripts, and plotting scripts that generate published figures. In this strict level of scrutiny, every artifact of the repository must have a public changelog, \eg a git commit history and transcript of AI prompts, which (almost) prove the chain of events from the creation of an artifact to the published results and interpretations. 

Crucially, a Level 4 result is traceable back to all raw observational data, includes the calibration steps, and describes every step of the data processing. Note that reaching Level 4 requires planning to reach it at the time a project is conceived. The publication that marked Level 2 must also include the full transcript log of the prompts to the AI agent(s) and a ``supervision log'' that explains what steps the researcher took that were independent of AI. A traceable project enables others to recreate the chain of events that led to a publication, but does {\it not} require different AI agents to produce the same results.

\clearpage
\subsection{Level 5: Regenerated\label{sec:proof-5}}

The penultimate step in the PROOF Ladder is the recreation of the results with at least one different AI agent and bringing this second analysis to Level 4. This regeneration should not consist of blindly feeding the prompts to another agent as the \vaibify paradigm is that the researcher is an active participant in the process and may have made manual changes to files and data. Instead, the researcher should use the publication from Level 2 as an implementation plan to collaborate with AI agents to create an entirely new pipeline that produces acceptably close results. 

The new results are exceedingly unlikely to match at the bit-level due to order-of-operations differences and instead need to be deemed consistent with the original publication. Thus, for a project to reach Level 5, the Level 2 publication should also include a list of the ranges of each output parameter that the researcher deems to be acceptable. While this latter point is a judgment call, publishing them exposes the choices to peer-review and enables Level 5 to be a quantitative assessment. A regenerated result has demonstrated that a researcher working with multiple AI agents can produce scientifically equivalent results, but does {\it not} require independent researchers to affirm that they have also regenerated the results with multiple independent AI agents.

\subsection{Level 6: Attested\label{sec:proof-6}} 

The highest level of the PROOF Ladder is all of the above, plus independent and publicly verifiable attestations by trusted entities that they regenerated the results with multiple agents and obtained results within the prescribed tolerances from the Level 2 publication. These concurrences would need to be independently verifiable in a public way, such as unit tests run on a public GitHub repository and/or cryptographic links to verified witnesses' identities, such as through ORCID keys that are linked to the DOIs of remote archives. Such attestations are, for example, enabled on Zenodo and via standalone packages such as \texttt{in-toto}~\citep{TorresArias19}. Multiple independent researchers affirming that they have ``regenerated'' a result is the strongest possible check on the role of AI coding agents in the production of a quantitative analysis.

Reaching Level 6, however, does {\it not} imply the results correctly describe the universe. It ``just'' means that some group of people (and/or trusted AI agents) have publicly and verifiably stated that a result has reached PROOF Level 5. There can still be corrupt observations, conceptual misunderstandings, and/or identical bugs introduced by the AI agents. Different scientists can of course also interpret results differently. The scientific method must march on to determine if the model actually matches the data through new, independent experiments. 

Level 6 is a high bar that requires significant effort from the community and will, in the near-term, probably only be applied to high-profile results, such as the claim of the discovery of life on another world or that we live in a broader multiverse. AI companies are now announcing high-profile scientific results prior to peer review and without releasing the data products that could enable researchers to check their veracity. Adopting the PROOF Ladder for scientific results and requiring declarations of the highest rung achieved could provide the scientific community (and the general public) immediate insight into how robust an AI-assisted or AI-generated claim is. 

\section{Discussion and Conclusions\label{sec:concl}}

The previous sections laid out a framework for determining how to reproducibly employ AI coding agents to write all the software of a data analysis project and described a methodology for measuring how well investigators understand the role of AI in their software development process. The keen reader, however, has noted that the previous sections are all just proposals at best or assertions at worst. Whether humans would spend fewer resources reaching Level 6 than by reaching Level 3 on their own remains an open question. More specifically, the question is whether verifying a result managed by AI agents is more difficult than a human-only analysis. We'll probably know the answer in a few years, but since AI coding agents represent a transformative approach to writing software, the only way to find the answer is to use AI agents to write our own code and evaluate how well they do.

The ideas and general philosophy of $\S\S$2--3 arose from the development of \vaibify, which is described in the Appendix. This package runs a web page that enables researchers to interact with multiple coding agents while they take their investigations from musings to publications to the permanent archival of byte-reproducible results. This code was built entirely by AI agents and followed the guidelines and goals outlined above. The researcher's analysis software runs inside a Docker container with a range of network security settings. \vaibify currently only supports research up to PROOF Level 3 (reproducible), but as described in the Appendix, reaching this level still requires a significant commitment by the researcher. \vaibify may allow the scientific community to assess the value of AI coding agents in the day-to-day research activities of scientists who analyze data sets.

Note that \vaibify itself was entirely ``vibe coded'' with the GUI written in a language I do not know (JavaScript), with the backend written in Python, which I do use. While the coding agents generated functional code, \vaibify's development required thousands of prompts, so current AI coding agents cannot trivially produce a tool of this sophistication. Nonetheless, \vaibify represents a complex scientific code that was generated entirely by AI agents that were partially steered by tens of thousands of tests and that, crucially, mostly works. Regardless, it's open source, so any researcher can download and tailor it to their own workflows by prompting agents and modifying the existing testing structure.

While many of the concepts described herein are similar to well-established procedures, others are new to the AI era and/or based on my own experience. Thus, the qualitative ideas presented here should be viewed as a possible, not definitive, guide to augmenting investigations with AI coding agents. Moreover, the AI revolution is evolving quickly and it is difficult to predict the long-term utility of the concepts presented in $\S\S$2--3. Nevertheless, coding agents represent a transformation in how scientists can generate and analyze data, so best practices need to emerge. Some scientists and software developers may reasonably never trust code written entirely by AI agents, but the concepts presented in this paper may lay the foundation for the scientific community to fully trust and embrace AI coding agents. As new LLMs are released and our collective  experience with them grows, the scientific community will learn how much cognition can be safely offloaded to AI while still ensuring that the rigor of the scientific method remains intact.

\vspace{1cm}
\noindent{\bf Acknowledgments}\\
I am grateful to Jess Birky, Francisco Villaescusa-Navarro, Eric Wolf, Jess Werk, Andrew Becker, Mike Brown, Alex Yates, Mike Way, and Scott Dally for conversations that clarified the concepts presented herein. I used Anthropic's \texttt{Claude Code} with the Sonnet 5.5, Opus 4.5--5.5 and Fable 5--5.1 models as well as OpenAI's \texttt{codex} with the ChatGPT 5.6 model to build \vaibify and drive the AI greenhouse example to PROOF Level 3. No LLMs were used in the writing of this manuscript. This effort was supported by NASA grants No. 80NSSC23K0261 and No. 80NSSC24K0856. 

\bibliography{here}

\appendix

\section{\vaibify\label{sec:vaibify}}

The main text laid out qualitative approaches for integrating AI coding agents into scientific research and evaluating their impact, but did not describe how scientists can actually apply and evaluate that assistance. To close this gap, I created \vaibify, a web application and data management system that can support multiple projects in a secure computing environment and monitor an investigation as it ascends the PROOF Ladder\footnote{Publicly available at https://github.com/RoryBarnes/vaibify}. This application creates a Docker container for AI agents to build code and allows users to prompt terminal-based coding agents, validate results through unit tests and visual inspection, decompose workflows into steps, track dependencies, require human confirmation of results, connect to external resources for archiving, and enable byte-identical reproductions of published experiments.  

\begin{figure*}
    \includegraphics[width=\textwidth]{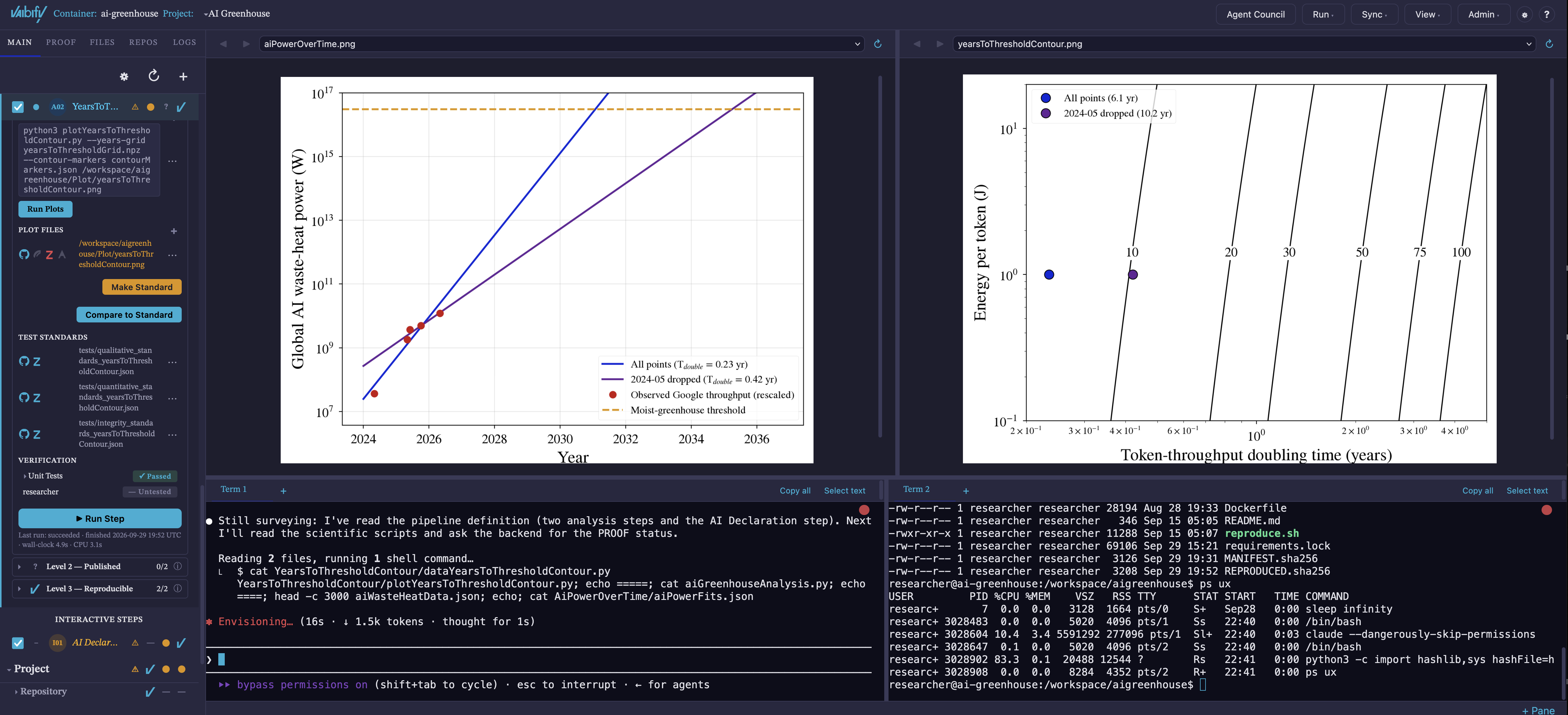}
    \caption{Screenshot of \texttt{vaibify}'s treatment of the example presented in App.~\ref{sec:aigreenhouse}. The very top shows high-level information about the project as well as pulldown menus to interact with the Project. The left-most side shows some information about the steps (currently expanding only step A02) as well as Project-wide requirements. Each step divides progress into the Level 1, 2, and 3 requirements, and includes icons that indicate the remote status (blue = synced with matching hashes; orange = not checked yet; red = hashes differ; dark grey = cannot be compared; light grey = not present on the remote). The Step's banner includes a checkbox for if the step should be run next time, a marker for how the last run ended, the Step's name and number, a space for any warnings, and finally three markers that display the current status of the Step in terms of the first three PROOF levels. The top row consists of viewing windows that display figures or ASCII text files, which  enable users to compare outcomes and ultimately verify that all steps are accurate. The lower row consists of terminal windows that enable the user to interact with a coding agent (\eg the left terminal shows \texttt{Claude Code} considering how to improve the project ) and/or work in the container directly (\eg the right terminal shows Linux output). Note that \texttt{Claude Code} is running in bypass-permissions mode because the containerized environment has tightly controlled access to host files.}
    \label{fig:gui}
\end{figure*}

\vaibify is designed for the general data analysis problem, but at this time, \vaibify can only confirm a project has reached Level 3. This Appendix describes how the package was created and how it addresses the topics in the main text. Fig.~\ref{fig:gui} shows a screenshot of the application while in the process of developing  the AI moist greenhouse example that concludes this Appendix.

\subsection{Fundamental Assumptions\label{sec:assumptions}}

The animating assumption of \vaibify is that visually verifying the outputs of AI-written software partnered with continuous integration (CI) is a faster path to good science than human-only written code and analysis. \vaibify's value proposition is that after an hour of setup, researchers can generate and reproduce results in a shorter amount of time than if they work without AI agents. This hypothesis will no doubt be tested in the coming years and \vaibify will likely evolve as the community gains more experience with AI agents.

I assume that researchers will use \vaibify to prompt agents to write code and hence an integrated development environment (IDE) is unnecessary. The package does contain a minimal text editor, but this feature is intended for small tweaks, like changing an input parameter. If necessary, a user can still view and modify the contents of the container via a \texttt{VS Code} IDE from the host machine. \vaibify thus enables researchers to use AI agents throughout the entire research process, from conception to tool building to complete numerical experiment, with the researcher only examining plots and data files to validate the results.

\vaibify assumes that a Linux file structure is ``ground truth'' and that output from a Linux command line is {\it always} accurate. While \vaibify may appear to only be a GUI application, all actions can be performed from the command line, including reaching PROOF Level 3. \vaibify is thus entirely scriptable, too, which could enable researchers to develop automatic procedures for any and all aspects of \vaibify's functionality. 

A fundamental requirement of \vaibify is that the dashboard view is always up-to-date so that researchers can immediately see if something has gone amiss. At regular intervals, \vaibify checks timestamps on files and checks that dependencies between steps are not broken. The polling interval's default value is 5 seconds.

\vaibify has a defensive security posture to both protect user data as well as minimize the attack surface. The container's user is unprivileged (uid 1000), with no \texttt{sudo} authorized. A container's security posture can be 1) full network isolation with no incoming or outgoing traffic of any kind, 2) only the few external services \vaibify and the running project need to perform the core functionality, 3) any public website with host access limited to \vaibify's port and no access to the local network (default), and 4) access to all public websites and all local services, including those unrelated to \vaibify (discouraged). Options 1 and 2 are primarily useful for scripting, but the last option should be used cautiously because an agent, or code it runs, could use any unauthenticated or vulnerable service on the host computer or its network. Zenodo and Overleaf credentials never enter the container, while (optional) GitHub tokens are read-only files in the container, never environment variables. Each container has a unique agent token to prevent an agent in one \vaibify container from modifying a different container. Commands sent to a container require a token naming that container, which is confirmed to exist at the moment of execution.

Finally, I note that I treated \vaibify as an example of how to develop AI-written code with the ideas of $\S$~\ref{sec:amenable} in mind. The project is monitored via git from the beginning and is provided a clear overarching instruction plan (AGENTS.md). I used deterministic methods as frequently as possible, often creating scripts after LLM prompting revealed that a feature could be handled by code. Unfortunately, most of the early prompts that generated \vaibify have been lost, so the prompt log is not included in the repository. I relentlessly requested architectural, security, and style reviews, which led to adopting CI tests for falsification, code style, and cybersecurity. I enforced human-readable function and variable names (Hungarian notation). Each merge is subjected to $>$16,000 unit tests that cover $\sim$95\% of the Python code, hundreds of browser and invariant tests, as well as $\sim3,000$ falsification tests via GitHub Actions. Unit tests are run on two Ubuntu and two macOS platforms and Python versions 3.9--3.14, with the status of these tests displayed on the \vaibify GitHub page. The functionality and philosophy behind the choices are fully documented online so that both humans and agents can learn why decisions were made. Thus, researchers who wish to build their own testing standards for their own scientific research codes can leverage the \vaibify methodology.

\subsection{AI Optimization\label{sec:addressing}}

In this subsection I describe how \vaibify specifically addresses each of the properties presented in $\S$~\ref{sec:amenable}. In some cases, I describe how \vaibify was itself developed as an example. 

\subsubsection{Quantifiability\label{sec:quantifiability}}

While some significant scientific progress has been made via qualitative means \citep[\eg][]{Faraday1839}, AI-assisted scientific experiments must be quantifiable. The stochastic nature of LLMs means that any qualitative results or judgments should be viewed skeptically. A script that generates numbers, however, can be rigorously evaluated, \eg confirming that energy is conserved. Thus, the hypothesis and methodology must be clearly articulated so that the AI-generated code produces the numbers that can falsify a hypothesis. 

\vaibify can do little in terms of confirming that a project produces quantifiable results as that aspect is largely the domain of the researcher. One important exception, however, is \vaibify's ``Agent Council'' that engages multiple agents to debate a researcher's question and is deemed successful  if all agents accept the plan, see $\S$~\ref{sec:contestatability}. Engaging the Council can, for example, help a researcher transform a vague, qualitative prompt like ``Make the results statistically robust'' into a specific, quantifiable prompt that is much more likely to generate a useful result: ``Use an affine-invariant MCMC sampler to infer the posterior distribution for all model parameters and require that the number of steps per walker is 50 times the auto-correlation length.'' Ultimately, however, it is up to the researcher to determine if the Agent Council has converged on a quantitative approach.

\subsubsection{Tractability\label{sec:tractability}}

As with all projects, an AI-assisted project must be able to be completed with the available resources and within a reasonable amount of time. A researcher may not know how tractable a project is at the beginning, though, especially since they are very unlikely to know how fast the AI-generated code will run. While it can be tempting to tackle much larger problems with AI, the process of verifying and publishing the results can be significant. Further muddying the tractability waters is that, at least for the time being, the abilities of AI agents are rapidly increasing; an intractable problem today may become tractable in a few months. So it may be worth it to start a groundbreaking project now in the hopes that more capable agents will be able to complete the project in a shorter amount of time in the near future. These assessments and decisions are mostly the purview of the researcher, but prompts to an AI agent could provide additional insight into the computational resources required for a project.

As with quantifiability, tractability is largely the purview of the researcher. \vaibify does include functionality to monitor CPU usage, allocate CPU(s) to a container, and keep track of how long individual steps take, all of which can help a user determine the tractability of a project. The Agent Council can again provide some guidance by estimating the CPU resources needed for a project. Because the Council is empowered to modify files, its members can test out various approaches to identify the optimal methodology, determine if sufficient resources are available, and suggest a minimally viable procedure.

\subsubsection{Containability\label{sec:containability}}

To protect a user's computer from damage by an AI coding agent, \vaibify creates Docker containers whose agents can only communicate with the host machine via a restricted, authenticated API and even then can only interact with explicitly exposed host machine files. Since Docker containers can create specific computing environments that promote reproducible research \citep{Boettinger15}, they are the ideal choice for \vaibify. Users can add software and data files to the container upon instantiation, access the web from inside the container, and direct coding agents to modify files. \vaibify containers are flexible, so if a user discovers that they forgot to add a repository or some other functionality when creating a \vaibify container, for example, they can easily add it afterwards from inside the container. They can also modify the \vaibify configuration file to ensure the forgotten package is a part of future builds.

\vaibify can be run in three modes: sandbox, toolkit, and project. Sandbox mode is for free-form exploration in a single repository -- researchers can create code, plots, and data as they choose. Toolkit mode is similar, but is intended for co-developing software packages that must work together. These modes only consider containability and no project being developed in these configurations can reach Level 1 of the PROOF Ladder. If the experimentation appears promising, the user can press a button and promote the work into a proper project.

\vaibify offers a container ``wizard'' that walks the user through the container build. This feature allows a large number of configurations, such as programming language support, AI harnesses, memory, CPU count, etc. If additional packages turn out to be required after the container is built, they are simple to add by following standard methods, \eg~\texttt{pip}.

While containability is a core feature, \vaibify can in fact be applied to a folder that exists in the host machine's own directory structure, \ie outside the container. This mode is not advised for most use cases as it can lead to permanent harm, but works well for machines dedicated to performing simulations. Since building the Docker environment can require more than 60 minutes, this mode sidesteps this cost and allows users to test out \vaibify's functionality before committing to creating a full Docker image. Additionally, \vaibify can connect to remote machines, so users can open multiple tabs to monitor all their projects, wherever they exist.

\subsubsection{Translatability\label{sec:translatability}}

There are many ways to generate translatable code, such as enforcing snake\_case or Hungarian notation, so \vaibify does not require users to follow any particular approach. Instead, users can declare rules in their own instruction files that will permanently guide the LLM in its naming conventions except for security concerns and the Agent Council. Note, however, that \vaibify's own source code is written with Hungarian notation.

The approach \vaibify's source code follows is roughly that of \cite{Martin08} and \cite{HuntThomas2020}, which emphasize human readability independent of comments, short ($<$30 lines) orthogonal single-purpose functions, minimal duplicate code, and modest, verifiable updates to the source code. The variable and function names follow the Hungarian notation approach with camel case names, \ie each variable/function begins with lowercase letter(s) that denote the cast or type and each word begins with a capital letter. I also require all function names to start with an ``f'' and include a verb. I required words to be spelled out instead of using acronyms or abbreviations so developers (human and agentic) can understand a function's purpose. The overall goal is for functions to read like paragraphs, lines of code to read like sentences, and variables to seem like characters in a story. In my experience, this approach maximizes readability, which is very valuable when reading AI-generated code.

\subsubsection{Localizability\label{sec:localizability}}

Localizing features and functionality to a small range of line numbers is an inherently difficult task for an arbitrary problem. \vaibify addresses it through its own AI harness, a set of natural language instructions that accompanies the code, see $\S$~\ref{sec:harness}. \vaibify's harness asks agents to try to limit functions to $<30$ lines, generate an architecture that is modular, and use a naming scheme that is easily greppable (features that also improve translatability). Note that the function length is not a strict rule. AI harnesses are fallible, however, so researchers should periodically prompt an agent to perform an architectural review to maximize localizability. \vaibify also includes deterministic methods to enforce localizability, see $\S$~\ref{sec:enforceability}.

\subsubsection{Contestability\label{sec:contestatability}}

As mentioned in $\S$~\ref{sec:quantifiability}, \vaibify includes an Agent Council in which the researcher can select 2--8 agents from different models, but not necessarily different providers, to debate a prompt and provide a consensus response. The protocols of the Council are as follows: 1) The researcher selects agents from available LLMs and designates one as the leader (the ``chairbot'') that will ultimately draft the final report. 2) Each agent receives its own ``shadow'' container each turn that enables it to run scripts, move files, etc., all without modifying the actual container. 3) The agents then independently create, review, and modify the response, ultimately providing the researcher with a (hopefully) robust response and/or plan of action and/or code.

Agent Councils are only permitted in a container because a) multiple agents are more likely to cause harm, and b) the Docker environment enables the quick (as short as a few seconds) creation of duplicate containers that the agents can use to test hypotheses, \ie they have write access. To convene a Council, the user presses a button and then follows the on-screen instructions to pick the agents' models, anoint the chairbot, provide the prompt for the council, and set modifiable parameters like the maximum number of rounds. If a council reaches that stage, then the researcher can resolve the issue, raise the number of rounds, or archive the results for future use.

Each agent receives the prompt, as well as the Council charter (\ie a separate harness included with \vaibify). During the first round, all agents independently examine and analyze their copy of the repository to discover the information required to answer the prompt.  Once all the members of the Agent Council have completed their analysis, they write an anonymous report that is distributed to all members for review and critique. At this point, the chairbot compiles the initial documents and their reviews into a single summary document and sends it back to the other agents on the Council for another review. They each either declare the document satisfactory or raise additional objections. The Council then iterates until 1) all agents agree to the response, 2) the maximum number of iterations is reached, or 3) any member decides it needs more information from the researcher.

Note that at each round, all agents begin with a new shadow container. This approach ensures no agent can be confused as to the contents of the true container. While this design should generate a more unbiased decision, Agent Councils are much more token-intensive than single-agent queries, so invoking one can become expensive. Thus, in many cases, developing a plan via the Agent Council may be the optimal approach for implementing a complex feature or developing a multi-step experiment. In addition, duplicating the container requires enough memory for all the shadows, so Agent Councils are not available for every repository. In some cases, omitting files from the Council's containers can be acceptable for large containers that could fill the machine's memory, so \vaibify allows the user to leave some files out of the members' shadow containers.

Two types of Agent Councils can be convened: plan and implementation. The former is for reviewing code and creating a roadmap for implementing a new feature or set of features. Implementation Councils plan a code change that is then written by the chairbot. During each council, users can monitor the progress, be prompted for clarifications, and ask questions to the chairbot. 

\subsubsection{Decomposability\label{sec:decomposability}}

As mentioned in $\S$~\ref{sec:amenable}, most scientific projects can be broken down into discrete, well-defined components and so are decomposable. \vaibify calls each component a ``Step'' and the full set a ``Project.'' The exact divisions between Steps are unique to each Project and therefore left to the researcher to define. Steps may be ``interactive,'' meaning the researcher must actively decide specific quantities, or ``automated,'' meaning the input data are present on-disk and a script can be run without user input. Each step can include a brief overview of the goal of the step, which can help LLMs understand the goal, write better scripts, and help other researchers understand the objective and how it fits into the larger goals of the Project. Explicitly decomposing the problem into steps can also make each accessible to the researcher, which could be critical for long and complex numerical experiments, see $\S$~\ref{sec:verifiability}. 

\subsubsection{Iterability\label{sec:iterability}}

Research projects rarely, if ever, work the first time they are tackled. Most research projects instead require iterating on the steps and the overall goals until the output provides insight. Iterating should be possible for both AI agents and the human researcher(s).

Breaking up a project into steps also facilitates iterating until all steps work properly. \vaibify is designed for AI agents to modify the steps based on the user's prompts until the researcher is satisfied. In general, \vaibify expects users will work with a terminal-based agent to revise and rerun steps until the results are accepted. A library of actions embedded in \vaibify enables coding agents to perform identical, deterministic actions that are also represented by buttons in the GUI.

\subsubsection{Parallelizability\label{sec:parallelizability}}

Modern computers have multiple cores and many current LLM harnesses support agent teams (or ``swarms''). Thus, projects in which an individual step can be parallelized across CPUs and/or GPUs are more efficient than serial codes. In addition, identifying individual tasks that can be independently tackled by an agent swarm will shrink the time for data analysis pipelines to finish. 

\vaibify only allows parallelizability via multi-core calculations. Steps must be run sequentially or individually. Thus, \vaibify's harness directs agents to create multiprocessing pools tailored to the container maximum CPU count. By default, a \vaibify container will be built with ($N-1$) CPUs, where $N$ is the total number on the host machine, \ie one processor is saved for the host machine. This number is configurable with the Container Wizard, see $\S$~\ref{sec:containability}. 

\subsubsection{Verifiability\label{sec:verifiability}}

To verify a component of a Project, \vaibify requires multiple independent tests to be satisfied for each level of the PROOF Ladder. Most of the step's requirements lie in Level 1, with higher levels' requirements mostly lying at the Project level, \ie the requirement applies to all steps equally. The status of each level's requirements is summarized in the GUI. The list of verification steps taken by \vaibify is listed below, with the PROOF level shown in parentheses, \eg L3 = PROOF Level 3.\\

\noindent{\bf Input Data (L1).} A list of files that contain the raw data that the step needs in order to complete. These files are tracked by the polling to ensure they are not corrupted. If no data are needed, \eg a script uses equations to produce output, then the user can declare that no input is needed.\\

\noindent{\bf Data Analysis Commands (L1).} The list of scripts/executables needed to complete the step, command line arguments, output files, etc. In principle, a Step can include an arbitrary number of commands, but usually it is best to make a step as small as possible. This piece is optional because steps may only generate figures.\\

\noindent{\bf Output Data (L1).} A list of the files that are expected to be generated from the Data Analysis Command(s). These data can be in an arbitrary format, but note that at Level 2 the names must be unique to satisfy Zenodo's policy. These files are monitored by the polling to ensure no unexpected changes to the file occur.\\

\noindent{\bf Plot Commands (L1).} The command(s) necessary to generate a figure. The format is similar to the Data Analysis Commands, and similarly is optional. These scripts are monitored by the polling.\\

\noindent{\bf Plot Files (L1).} The output file(s) of the Plot Commands. The researcher can create different file types (\eg png and pdf) to examine how they change after modifying a script or input data. \vaibify allows researchers to declare a plot to be the standard that is saved and can be compared to newer versions as they become available. This feature can be helpful if code needs to be restructured, but the result is expected to be preserved.\\

\noindent{\bf Test Standards (L1).} A file that contains the quantitative output expected from a Data Analysis command. These files are used for unit testing as described next. These files are optional, but note that \vaibify includes functionality to easily generate unit tests. If they exist, these test files are monitored for changes.\\

\noindent{\bf Integrity Tests (L1).} \cite{Chen25} notes that LLMs can sometimes fail to follow the format of a given file, which can cause an error that prevents any verification that an output value matches a standard. \vaibify thus runs a per-step unit test of the files that confirms they are not corrupt.\\ 

\noindent{\bf Qualitative Tests (L1).} In some instances, scripts will output descriptions or classifications of the results that are not amenable to numerical analysis. Qualitative tests verify that these descriptions of the output remain constant.\\

\noindent{\bf Quantitative Tests (L1).} These unit tests confirm that each numerical output from the Data Analysis Command(s), such as specific values and/or median values of an array, are within a user-specified tolerance to the standards. Constructing an algorithm that can deterministically generate these unit tests for arbitrary data structure is a very challenging problem, so \vaibify ships with a simplified approach based on suggestions from \cite{OberkampfRoy10} and \cite{Vehtari21}, assuming that users will augment this foundation to meet their specific needs if necessary, \ie asking an agent. If the deterministic approach doesn't work, \vaibify falls back to an LLM to generate the test scripts.\\

\noindent{\bf Researcher Approval (L1).} \vaibify requires the user to confirm the results are accurate at every step in the pipeline. This step ensures that a human is in the loop and enables accountability for not just the entire research project, but for each pipeline component. There are no restrictions on when or why a researcher decides a step is accurate. This quantization of accountability could potentially help ensure that future, complicated (and high-profile) results produced by hundreds or more individual steps can be understood and trusted by the community. This requirement of human verification distinguishes the \texttt{vaibify} approach from other AI-assisted research tools, such as \texttt{Denario} \citep{VillaescusaNavarro25}\footnote{Publicly available at https://github.com/AstroPilot-AI/Denario.} and \texttt{autoresearch}\footnote{Publicly available at https://github.com/karpathy/autoresearch}, which are intended for fully autonomous investigations. \\

\noindent{\bf Dependencies (L1).} If a step depends on the results of an earlier step in the pipeline, then that previous step must be fully verified for the current step to also be considered fully verified. Moreover, the {\it timing} and {\it content} of the files created in the steps also matter: a file used in a subsequent step must have been created before the analysis in the subsequent step was started or, if not, then the bytes of that file must match the bytes when the dependency was last verified. \vaibify monitors file timestamps and SHA-256 hashes to ensure that a verified step isn't inadvertently modified later by either an agent or a human. Should a previously verified dependent step be modified, the downstream steps' dependencies flip to failed and the researcher is notified in the GUI. Similarly, if a user manually changes a file, \vaibify catches that change and updates the GUI accordingly. This constant monitoring (``polling'') ensures that the current status of the workflow is self-consistent and presented to the researcher.

Automatically detecting all dependencies in a general workflow is another intrinsically difficult challenge. \vaibify auto-detects explicit dependencies via scanning code for read functions to identify files created by other steps across 11 coding languages (Python, R, C/C++, Fortran, Rust, JavaScript/TypeScript, Perl, shell, Julia, MATLAB and Go), which is deterministic and discovers many clear dependencies between steps. However, some steps can implicitly depend on others in subtle ways that are impossible to detect automatically. Thus, \vaibify also allows researchers to reject automatically detected dependencies and manually assign dependencies to ensure that any step with a subtle or invisible dependency is properly tracked.\\

\noindent{\bf GitHub Mirroring (L2).} The output of the step matches the files that are stored on the remote GitHub repository. This validation is performed via SHA-256 hashes that determine if the two files are byte-identical. This check, and others that require communication with remote repositories, is performed when opening a Project, but is subsequently only checked once every 6-24 hours due to the potentially large latency in connecting with the remotes. Users can, however, verify a given file or the whole repository at any time.\\

\noindent{\bf Zenodo Mirroring (L2).} The same as for GitHub, but for the Zenodo archive. Note, however, that a Zenodo archive consists of the set of files used to generate a publication, \eg output files, unit tests, inputs, \vaibify-specific files, the AI declaration and the PROOF Level 3 attestation.\\

\noindent{\bf Overleaf Mirroring (L2 -- optional).}  \vaibify monitors the local and remote figures on Overleaf to ensure they have not deviated.\\

\noindent{\bf arXiv Mirroring (L2 -- optional).}  The same, but for arXiv. In this case, the remote record is downloaded and the individual figures are confirmed to be byte-identical. Note that this check only works {\it after} publication.\\

\noindent{\bf Outputs in Manifest (L3).} Confirmation that the files in the step are included in the list of files/data that must be reproduced. Additionally, random seed declaration, additional binaries, and CPU settings are also recorded.\\

\noindent{\bf Scripts Match Manifest (L3).} Confirmation that the scripts for the step are included in the reproduction manifest and are bit-identical.\\

The \vaibify GUI displays the real-time status of all these properties on the left of the application, see Fig.~\ref{fig:gui}. Users can run commands ``manually'' by pushing virtual buttons, or ask their AI agent to do so (see $\S$~\ref{sec:harness}). When a piece fails the requirement, warning glyphs are displayed with error messages that explain the problem. 

Note that several of these features resemble those in existing packages such as the workflow management system \texttt{Snakemake}~\citep{Molder21}. This package was designed for reproducibility in the pre-AI era and therefore does not include functionality for human validation, remote mirroring, and declarations of AI use. The distinguishing feature of \vaibify is that it assumes AI agents write the code, which elevates these additional features to fundamental requirements.

\subsubsection{Observability\label{sec:observability}}

\vaibify addresses observability in multiple ways. The dashboard format, in which issues with a project's dependencies and outputs are quickly visualized, allows researchers to see failure modes plainly. \vaibify also provides a deterministic method to identify where the code may exit without a clear explanation (although catching {\it every} possible failure mode is probably impossible). 

\subsubsection{Reproducibility\label{sec:reproducibility}}

\vaibify addresses reproducibility and archiving by integrating with GitHub, Overleaf, arXiv, and Zenodo. The application allows the user to select files to store remotely and securely stores credentials in order to monitor SHA-256 hashes of both local and remote files. These hashes allow any other scientist to rerun the results and confirm the local files on disk are the same as those in the published paper and on GitHub.

\subsubsection{Falsifiability\label{sec:falsifiability}}

The \vaibify AI harness (see $\S$~\ref{sec:harness}) is instructed to write code in a falsifiable style as much as possible. As stated in $\S$~\ref{sec:amenable}, achieving 100\% falsifiable code is almost assuredly unrealizable, but directing the agent to develop in this way at least increases the fraction of falsifiable commands. \vaibify itself was developed in this manner with over 3,000 falsification tests, which often revealed critical, but difficult to spot, mistakes in design and implementation by intentionally breaking code.

\subsubsection{Enforceability\label{sec:enforceability}}

\vaibify does not include functionality to enforce design decisions as they are up to the researcher. The \vaibify repository, however, does enforce these decisions via CI and these tests can serve as an example for investigators wishing to enforce their own code's design. The \vaibify repository hosts three classes of style requirements, which are called ``invariants.'' Architectural invariants enforce decisions about how the code interacts with the file tree, how the management functions connect to the browser functions, as well as numerous other aspects of how the code is built and tested. Security invariants address how the package handles multiple sessions, agents, and containers and how the web browser interacts with multiple sessions, agents, and containers. Style invariants enforce source code decisions such as Hungarian notation and human-readable variables (translatability).

\subsubsection{Archivability\label{sec:archivability}}

The output of any investigation should be easily transferable to permanent storage with tests that ensure that the local data files match the remote data files. This process includes placing data in standard repositories like Zenodo, but also connections to version control systems like GitHub, and manuscript preparation tools like Overleaf. The concern again is that an agent could silently modify a local file that has already been placed in a remote repository for long-term storage, or vice versa.  

Archiving is supported with API calls to Zenodo, a well-known long-term storage site for digital artifacts. Users can select files for long-term storage, build their Zenodo repository through its ``sandbox'' mode, and upload the final version into regular repositories for permanent storage. The UI monitors the hashes of these files, too, so should any file diverge, the GUI alerts the user.

\subsubsection{Replayability\label{sec:replayability}}

Although replayability is a Level 4 requirement, which is beyond the scope of \vaibify v1.0, some functionality still exists. This decision is motivated by the fact that in order to reach Level 4, the transcript of AI prompts must be recorded {\it prior to Level 1}. As prompts can contain personal and/or confidential information, \vaibify enables partial redaction of the prompt transcripts, but users should still review transcripts for private information before committing the prompt logs. Users may also store a hash of their private AI instructions file (\eg AGENTS.md) if they don't want to release them with a publication.

\subsection{Ascending the PROOF Ladder\label{sec:ascend}}

Users can start their investigation in a sandbox or toolkit and promote it to a Project, or they can just enter Project mode from the beginning and start creating steps. Project mode is for numerical experiments that are expected to become part of a document, \ie they have clearly defined steps that generate data products that require careful monitoring. As a project matures, \vaibify provides details on how to achieve the next PROOF level and notifies researchers when a Project reaches it. In this subsection I provide an overview of how to drive a Project to PROOF Level 3, but defer many details to the online documentation. 

\subsubsection{Level 1: Self-Consistent}

A Project at PROOF Level 1 is ``self-consistent,'' meaning all dependencies are verified and all defined tests for each step are passing. Table \ref{tab:L1} lists all the requirements for a Project to reach Level 1. The scope may be either ``Project,'' meaning it applies to the entire experiment, or ``Step,'' meaning the requirement must be met for each individual component of the experiment.

\begin{table}
\centering
\caption{PROOF Level 1 Requirements}
\begin{tabular}{ll}
\hline\hline
Requirement & Scope\\
\hline
Git enabled & Project\\
Input data declared & Step\\
Unit tests pass & Step\\
Your sign-off recorded & Step\\
Nothing changed since verification & Step\\
Cross-step reads declared or dismissed & Step\\
\hline
\end{tabular}
\label{tab:L1}
\end{table}

\subsubsection{Level 2: Published}

All requirements beyond Level 1 are primarily about disseminating and reproducing the results, so the focus shifts dramatically from science to bookkeeping. A project at Level 2 has been, or will be, submitted for publication or sent to other stakeholders. \vaibify assumes the results are published online and facilitates synchronization with the latex manuscript generation platform Overleaf, with arXiv remaining optional. In addition, mirroring with GitHub and Zenodo is enabled, completing the process of manuscript preparation, software curation, and artifact archival.

Table \ref{tab:L2} presents the specific requirements for a \vaibify project to reach Level 2. The requirements divide cleanly into two camps: those that are related to a remote repository, and those that can be handled locally. The personal AI configuration consists of the researcher's own instructions, skills, hooks, and memory files.  Note that the optional supervision and prompt logs are intended for reaching Level 4, along with the tolerances required to reach Level 5.

\begin{table}
\centering
\caption{PROOF Level 2 Requirements}
\begin{tabular}{llll}
\hline\hline
Requirement & Scope & Location & Required\\
\hline
GitHub mirror & Project & Remote & yes\\
Published files match the GitHub mirror & Step & Remote & yes\\
GitHub repository is public & Project & Remote & yes\\
Zenodo archive & Project & Remote & yes\\
Published files match the Zenodo deposit & Step & Remote & yes\\
Zenodo archive is public & Project & Remote & yes\\
File names unique across the deposit & Step & Remote & yes\\
AI declaration signed off (and current) & Step & Local & yes\\
AI models & Project & Local & yes\\
Personal AI Configuration & Project & Local & yes\\
Environment archive (answered; declining passes) & Project & Local & yes\\
Overleaf manuscript / figures frozen in Overleaf & Project / Step & Remote & optional\\
arXiv submission & Project & Remote & optional\\
Prompt Record & Project & Local & optional\\
Supervised mode & Project & Local & optional\\
\hline
\end{tabular}
\label{tab:L2}
\end{table}

\subsubsection{Level 3: Reproducible}

Reaching Level 3 means that the results can be reproduced -- exactly reproduced in many cases -- and that the instructions to do so have been articulated. Level 3 requires a comprehensive accounting of the files, software packages, computing environment, and Docker configuration to be published and monitored along with their SHA-256 hashes. Additionally, a script to reproduce the results, which \vaibify can generate, must be published. \vaibify requires that a result at Level 3 has been generated within a containerized environment to avoid dependency and environmental inconsistencies with a researcher's machine.

One key aspect of Level 3 is the requirement to archive the Docker environment to enable exact reproducibility. Docker slowly changes the environment with time, and while newer versions will be IEEE compliant, that does not necessarily mean byte-identical. For example, transcendental functions like \texttt{tan} can deviate in the final decimal place. Therefore, to enable byte-reproducibility, the list of environmental conditions (the ``environmental snapshot'') is insufficient: The environment must be archived in a public repository. \vaibify therefore enables users to either archive their current environment to Zenodo, or reference an existing archive with the exact same architecture. AI agents are not required to be placed in the archive as the results should be reproducible without them. Note that \vaibify verifies the archive contains the same pinned Docker image for the same CPU architecture. This approach may allow researchers to exactly reproduce research for over 20 years, which is the current funding plan for CERN. Note that 10 years of reproducibility is a surprisingly difficult goal to achieve \citep{Perkel20}.

Level 3 requires an attestation from the researcher that they have rebuilt the results from the reproduction script. This reproduction may succeed on other machines provided the Project's author has prioritized reproduction over speed. \vaibify enables users to perform this last step by creating a copy of the container, similar to the shadow containers used by Agent Councils (see $\S$~\ref{sec:contestatability}), that runs the project's steps to enable researchers to test if the project is byte-reproducible. \vaibify distinguishes byte-identical reproduction from reproduction that passes the published tolerances but possesses different output bytes. Note that in this case, the outputs of interactive steps are treated as inputs to other steps.

Finally, Level 3 requires the user to prioritize either speed of computation or reproducibility. Modern CPUs have numerous settings that can increase speed, but may result in differences in the numerical output. Rather than force a user into a slower processor mode, \vaibify allows users to select faster speed for their projects. 

Crucially, the order of the final steps to Level 3 cannot be random. For example, the reproduction script must be generated before the SHA-256 hashes are computed and stored so that its own hash is recorded. \vaibify includes features that guide the researcher through these final steps to prevent issues such as rebuilding a Zenodo archive or re-running the entire pipeline to attest its reproducibility. 

\begin{table}
\centering
\caption{PROOF Level 3 Requirements}
\begin{tabular}{llc}
\hline\hline
Requirement & Status & Order\\
\hline
Software declared & Declared or waived & --\\
Dockerfile pinned & Base image digest-pinned & --\\
Environment snapshot & Image digest recorded & --\\
Determinism declared & Answered; random numbers seeded & 1\\
Reproduce script & Exists, pinned in manifest & 1\\
Reproduce script is current & Matches generator & 1\\
Environment archived & Matches digest and architecture & 1\\
Dependency lock & Exists, hashed & 2\\
Manifest complete & Covers declared files and envelope & 2\\
Manifest matches the files & Hashes match current bytes & 2\\
Rebuild attestation & Passed against current manifest & 3\\
Envelope published (GitHub) & Matches GitHub & 3\\
Envelope archived (Zenodo) & Envelope and attestation in archive & 4\\
Archives are permanent & No sandbox deposit & 1 / 4$^a$\\
Image on a registry (optional) & Pullable & --\\
\hline
\end{tabular}
\label{tab:L3}
\\$^a$ The environment archive at 1, the project deposit at 4
\end{table}

\subsection{Code Integrity\label{sec:maintenance}}

As mentioned above, the \vaibify source code was written entirely by AI agents that were directed by the author. The complexity requires a large codebase: The browser interface consists of nearly 45,000 non-comment lines of JavaScript code, the core resources consist of over 160,000 lines of Python code, and the test suite is $\sim350,000$ lines of Python code. The approximately 2:1 ratio of test-code to production-code arises from the strict enforcement of testing the entire codebase. The value of this ratio is difficult to quantify, but this choice has clearly led to the successful creation of a $\sim$200,000 line package that is fully verifiable and without the author ever opening a single source code file.

The true benefit of these tests comes from leveraging GitHub Actions, which are free for open source projects, to treat the tests as CI checks.  GitHub Actions are specifically designed for CI and block merges for which any single test fails. Moreover, GitHub Actions allow for multiple permutations of operating systems and language versions. \vaibify tests against two Ubuntu versions, two macOS versions, and Python 3.9–-3.14. Pull requests run the complete Linux unit-test matrix and the oldest and newest supported Python versions on macOS; the complete macOS matrix runs nightly. Separate checks exercise browser behavior, security boundaries, architectural and style invariants, and falsification tests.

The first class of CI checks are the unit tests. These are tests of individual outcomes of a code execution path. For a code like \vaibify, these can span a wide range of functionalities, but for scientific software, they will often just be numbers that represent the result of a calculation \citep[see \eg][]{Barnes20}. \vaibify's unit tests only cover $\sim$95\% of the lines of code because achieving 100\% is effectively impossible without creating tests for each failure mode, which can create a testing process that takes hours. \vaibify's unit tests require about 30 minutes per OS/Python-version combination.

The next CI class is the invariant tests, of which there are three: architectural, security, and style. These tests focus on syntax and so do not require the full range of OS/Python-version combinations. The security tests are performed on Ubuntu 24.04 and macOS 26 with Python versions 3.9 and 3.14 to check the bounds. Architectural invariants run with unit tests but have their own dedicated CI check. The style checks are run on only one machine. There are currently $\sim85$, $\sim230$, and $\sim35$ architectural, security, and style tests, respectively.

Next are the browser tests in which the Python package \texttt{playwright}\footnote{Publicly available at https://playwright.dev/python/} performs mock interactions with the GUI to confirm that actions result in the expected outcomes. Not all actions are reproducible in \texttt{playwright}, such as GitHub syncs, but these tests ensure that a significant number of UI features do not break during development. As most scientific software packages do not include a GUI, this class of CI check will not likely be relevant to most projects in the near-term. However, as \vaibify demonstrates, sophisticated GUI packages are now accessible to scientists with minimal web development skills, so future codes may not be limited to the command line. \vaibify currently performs over 500 browser tests.

\vaibify also checks a custom set of falsification tests that ensure changing some functionality actually does break unit tests. As mentioned above, testing these ``mutations'' can take a long time, over 60 minutes for \vaibify even though they are broken up and run in parallel. Despite this relatively long duration, the falsification testing has regularly turned up errors that were not captured by the unit or invariant tests. There are currently $>3,100$ falsification tests.

Finally, \vaibify includes a handful of small sanity checks to ensure smooth behavior that are run at varying cadences. At each pull request, the agent documentation files associated with \vaibify, such as the paths to AGENTS.md and SKILL.md files, are confirmed to exist in the correct locations. Also on each pull request, the ability of the code to remotely ssh into the same machine is tested. Once per day, a minimal container is constructed that checks repository/workflow discovery, missing-file and directory behavior, atomic copy/rename operations, the unprivileged container-user declaration, file mtime and SHA-256 reads, Python execution, rerun-manifest verification, detection of changed outputs or rewritten manifests, and refusal to guess when multiple workflows exist. Finally, once per week, a fresh container is built from scratch to ensure this core feature has not been corrupted. In total, 16 automatic testing suites monitor the code for errors and inconsistencies. 

Once all tests have passed and the new version is merged into the main branch, GitHub Actions are also employed to build the documentation and every new declared version is released through \texttt{pip}. Given LLMs' ability to generate the Actions' instructions, these workflows were trivial to implement. The \vaibify GitHub page displays the current status of these CI actions.

This long list of maintenance procedures monitors code structure, style, and output, but does come with significant time and resource costs. Running the full suite on a local machine can require over an hour, often only finding a handful of issues per run. There are two reasons, however, why this trade-off is worth it. First, these tests prevent agents from inadvertently breaking existing functionality that was outside the context of its immediate prompt. Second, the tests are deterministic and hence do not invoke LLMs to read code and diagnose bugs. In other words, they are more reliable {\it and} save tokens. When bugs are discovered, the LLMs do spend tokens to fix the problem, but they know immediately where the problem is and how it failed, which can significantly speed up the time for the fix to be implemented. While an hour may seem like a long time to wait for the testing suite to finish, it does offer the time for other tasks that are typical for a scientist, such as preparing manuscripts, proposals, lectures, etc.

Finally, I note that these tests should enable more robust contributions from the community. The extensive documentation and design decisions allow developers to employ agents to modify the code efficiently. Moreover, researchers interested in integrating AI coding agents into their own research can follow the \vaibify examples to build and maintain their own codes.

\subsection{The vaibify AI Harness\label{sec:harness}}

As mentioned previously, \vaibify ships with its own set of instructions to help researchers maximize their return from AI coding agents. These harnesses arrive at multiple levels in the architecture and address different needs. The first is the \vaibify agent instruction file that is included with every container build and that can provide some top-level instructions for AI agents operating inside a \vaibify container. The second type of harness is the ``Agent Council'' described in $\S$~\ref{sec:contestatability}, which supersedes the user's prompt to ensure that members adhere to the rules of a council. Finally, numerous tasks are described as ``skills'' inside the \vaibify container.

I refined these instructions while developing my own codes/projects, but have kept them general. These harnesses are easily modifiable for specific tasks, but they do include a statement that the user's own instructions supersede the general instructions. Thus, a new user should not have to do anything more than insert their own file in the appropriate directory in a container. 

Several features of the instruction file are noteworthy. First, it includes an explanation of \vaibify, the rules of the container (if applicable), and a description of the PROOF Ladder. Second, it explains how to connect host functions to container commands, \ie the prompt describes how and when to use deterministic functions to perform actions that the researcher requests. For example, a researcher could either push a button to build unit tests, or ask an AI agent to do so. In the latter case, the instruction file {\it should} direct an agent to execute built-in functions that run the unit test build script rather than construct a unit test itself. Other requests direct the agent to a skill file that contains detailed and verified instructions that enable a specific, commonly used directive. Third, the prompt implores the agent to not assume quantitative results, but to build scripts that deterministically compute the values in question. While modern LLMs tend to now include instructions that encourage these behaviors, the \vaibify harness may be more explicit. While this approach can take more time, researchers need facts, not assumptions.

\subsection{Example Project: The AI-Induced Moist Greenhouse\label{sec:aigreenhouse}}

In this subsection, I present a simple example to demonstrate how \vaibify works in practice. The research question I ask here is ``When will AI adoption make Earth uninhabitable?'' The true answer of course is a complex mixture of physical, political, and social factors, so here I reduce it to a very simple physical model based on current trends in AI adoption and data center construction. The project can be reduced to just two steps (plus the AI Declaration) that can generate figures for easy interpretation.

First, some context: A moist greenhouse occurs when a planet's tropopause temperature becomes warm enough that water vapor does not condense there, which allows water molecules to penetrate the stratosphere where they can be destroyed by UV radiation. The liberated hydrogen, being the lightest element, can then easily achieve escape velocity via absorbing more photons and/or colliding with other molecules. Once all the water is destroyed and their constituent H atoms gone, the planet is no longer habitable. Careful calculations over the previous 5 decades have suggested that the tropopause temperature reaches this critical value when the thermal (infrared) energy flux emitted by the planetary surface reaches approximately 300 W/m$^2~(\equiv F_{MG})$. Currently Earth's surface is emitting about 240 W/m$^2~(\equiv F_0)$ in the IR, so once AI data centers contribute an additional 60 W/m$^2$, Earth will tip into a moist greenhouse and begin desiccating. In other words, we can reframe the earlier question as ``When will the waste energy from data centers generate 60 W/m$^2~(\equiv F_{crit})$ averaged over the Earth's surface?'' Note that the first form of the question was not quantifiable -- it was too general -- but the second form is more amenable to AI agent assistance.

While 60 W/m$^2$ may initially seem like a relatively small number, averaged over the surface of the Earth, it implies a huge amount of power: 30,000 TW. This amount is about 1 million times more power than is currently used by data centers across the globe, which is estimated to be about 50 GW \citep{Takci25}. It would thus appear that an AI-induced moist greenhouse is a distant prospect.

Nevertheless, I press on and frame the question in terms of tokens, the base unit of an LLM. In particular, I seek to calculate how many tokens must be processed globally per second to trigger the moist greenhouse, assuming the current energy requirements to process 1 token. The latter value depends on the hardware in question and appears to range from 0.1 -- 20 J/token, with most modern hardware below 1 J. I adopt 1 J/token~($\equiv E_{token})$ as a fiducial value and assume that all the energy goes into the environment. Thus, the current rate of token processing, $n$, is $5 \times 10^{10}$ tokens/second, and the rate needed to trigger a moist greenhouse is $3 \times 10^{16}$ tokens/second.

Next, I must speculate how the global token use is evolving from limited data. While it is clear that energy consumption for data centers is rising rapidly \citep{Takci25}, most companies do not report their token throughput, so this estimate will necessarily be crude. As a starting point, public statements from Google about their token use over time can be extrapolated to global token use. In May 2024, Google reported $n_{Google} = 9.7 \times 10^{12}$ tokens/month = 3.6 Mtokens/sec (Mtoken = million tokens); in May 2025 the number rose to 179.2 Mtokens/sec\footnote{https://blog.google/innovation-and-ai/sundar-pichai-io-2026\label{fn:pichai}}; 485 Mtokens/sec in October 2025\footnote{https://the-decoder.com/google-boasts-1-3-quadrillion-tokens-each-month-but-the-figure-is-mostly-window-dressing/}; and 1200 Mtokens/sec in May of 2026\footref{fn:pichai}. From these numbers we must estimate the growth rate, which I will assume follows a Moore's Law style functional form with a doubling timescale $T_{double}$, \ie after a time $T_{double}$, $n$ grows to $2n$. Finally, we must estimate the fraction of the global token throughput managed by Google, $f_{Google}$. With no available data on the topic, I choose the fiducial value to be 10\%, primarily for computational ease.

\begin{figure*}
    \includegraphics[width=\textwidth]{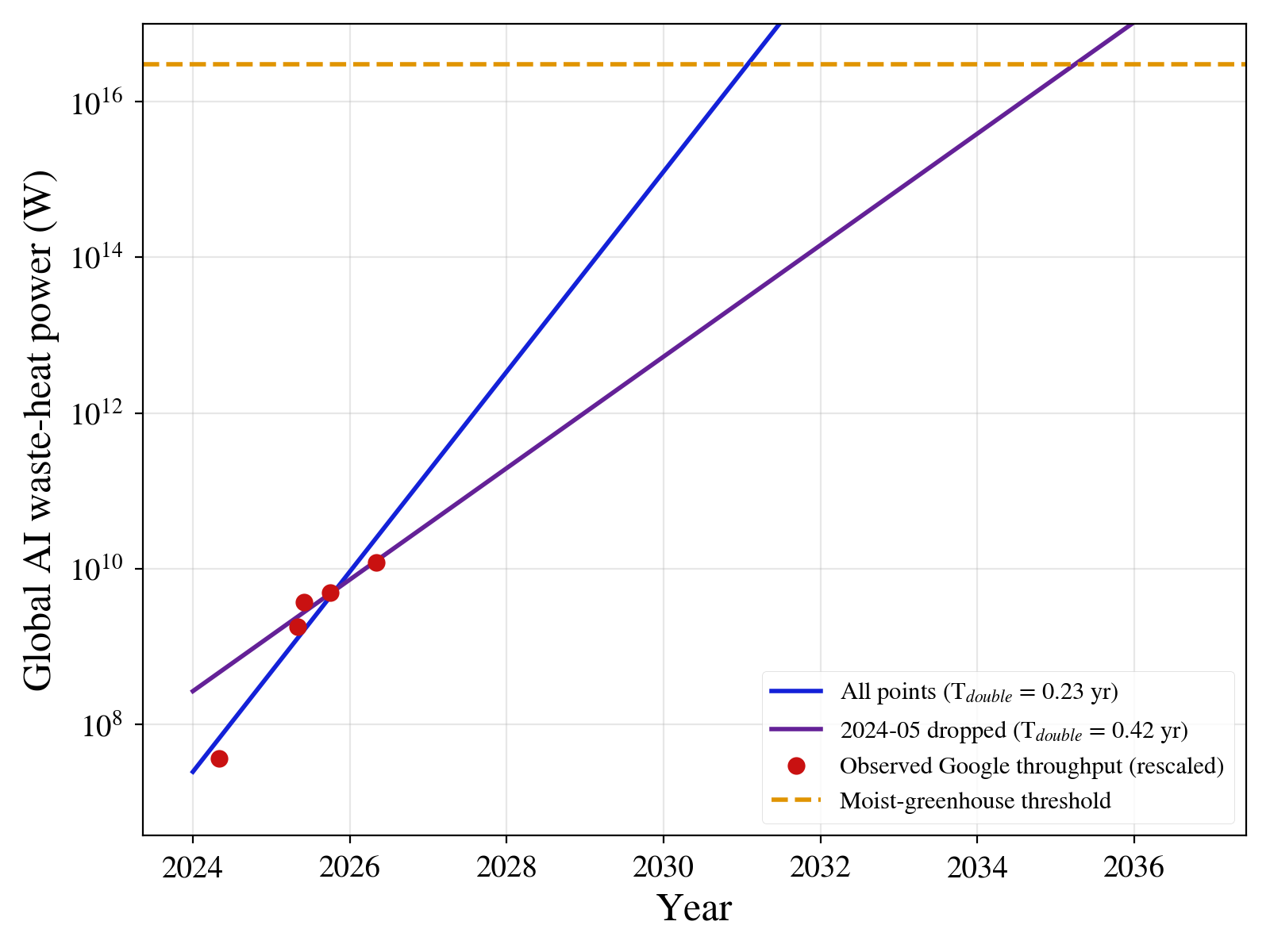}
    \caption{Power required to process all tokens as a function of time. The red points represent Google's stated token usage since May 2024 and assuming each token requires 1 J of energy and that Google's reported throughputs equal 10\% of the global rate. The blue and purple curves represent a fit to all the data (blue) and all but the earliest datum (purple). The dashed orange line represents the power required to drive the Earth's atmosphere into a moist greenhouse.}
    \label{fig:power}
\end{figure*}

From these assumptions, we can compute the time until AI data centers produce 30,000 TW of power. First, Fig.~\ref{fig:power} shows the time to reach this limit based on two doubling time estimates. In blue, all the data points are included and the Earth reaches the moist greenhouse limit in early 2031. However, the data also suggest $T_{double}$ is increasing with time and that the token throughput is {\it not} following Moore's Law style growth. For example, if we throw out the May 2024 data point, then the moist greenhouse limit is not achieved until 2035. 

\begin{figure*}
    \includegraphics[width=\textwidth]{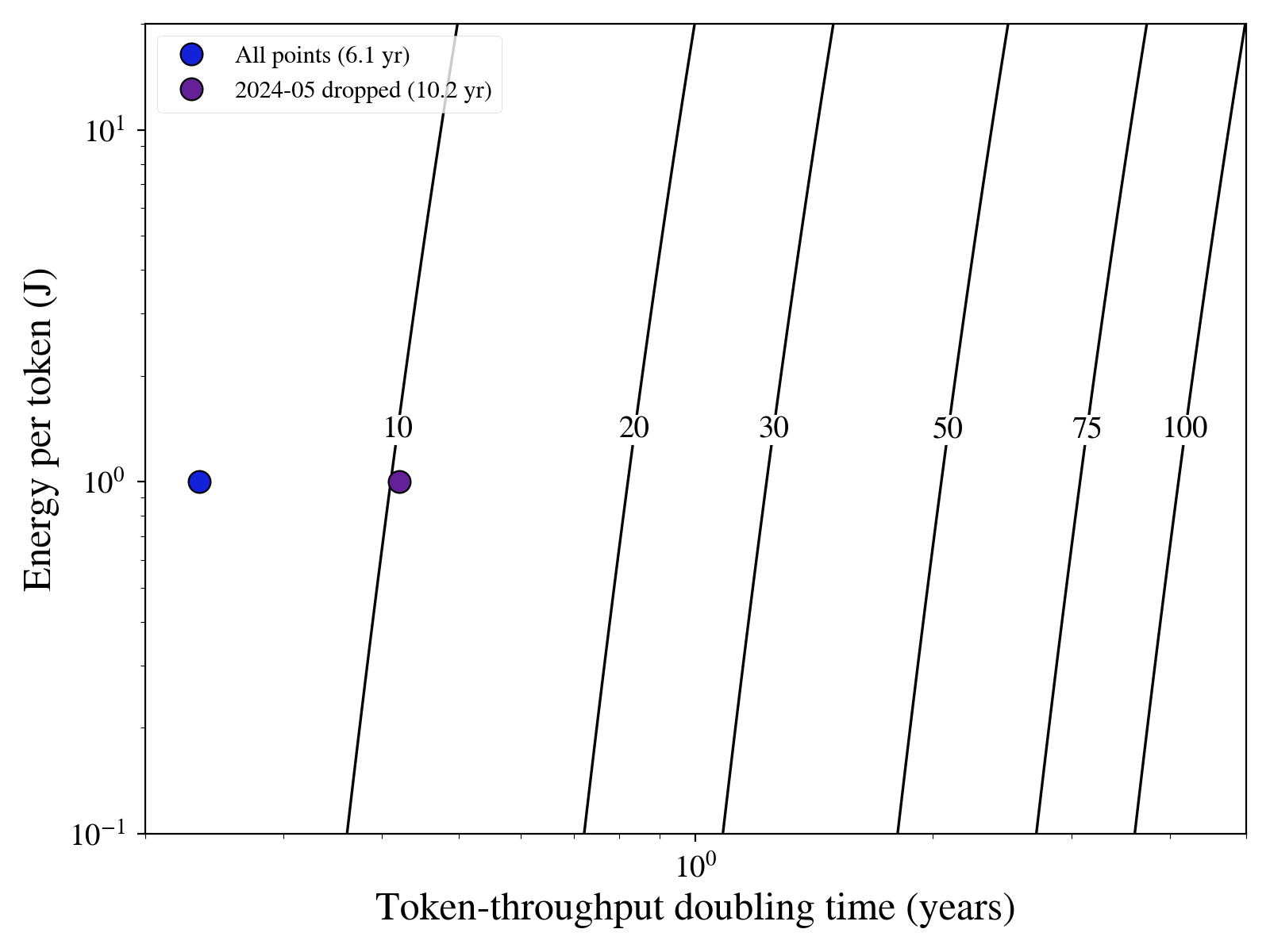}
    \caption{Time in years after January 2025 until token processing triggers a moist greenhouse, using the same assumptions as Fig.~\ref{fig:power}. The blue and purple dots represent the predictions from Fig.~\ref{fig:power}.}
    \label{fig:contour}
\end{figure*}

Fig.~\ref{fig:contour} shows the time from 2025 until the moist greenhouse limit is reached for a range of assumptions in energy/token and the doubling time, but with everything else held constant. Should the doubling time increase to several years and the energy/token drop, then the threshold is not crossed for over 100 years. This latter prediction is in line with others that have explored humanity's impact on the climate and the thermodynamic limits to industrial capacity \citep{Murphy21}.

This analysis was generated with AI assistance via \vaibify, both in terms of researching the core numbers as well as in generating the plots. After collecting the raw data, the figure creation was trivial. While the results are startling (more below), one need only look at the raw data (red points in Fig.~\ref{fig:power}) to see that the doubling times are reasonable and that the analysis, modulo the assumptions, is correct. I started this research out in sandbox mode, then once I was satisfied with the style of the figures, I converted it to a Project via a button push, and then generated the unit tests with two more button pushes. I then synced the results to GitHub and Zenodo with two more button pushes. This result on its own is obviously not publishable, but the GitHub repository is available online for anyone interested in exploring these calculations and/or testing out \vaibify\footnote{http://github.com/RoryBarnes/aigreenhouse}. Note that I have carried this investigation all the way to PROOF Level 3\footnote{Environmental archive available at https://zenodo.org/records/23897220.},\footnote{Project archive available at https://zenodo.org/records/23900492.}, so anybody can check that every byte of their reproduction is identical to mine. 

Before concluding this section, I need to emphasize that {\it the previous analysis should} {\bf not} {\it be construed to imply that the Earth will become uninhabitable in 5 years.} Instead, it more likely indicates that new innovations are necessary for the global token processing rate to continue to grow after 2030. Other considerations are also noteworthy, such as reaching the moist greenhouse would require approximately 60 million modern data centers that, with a typical size of $\sim10^4$ m$^2$, would cover 0.1\% of the Earth's surface! These back of the envelope calculations suggest that for the AI revolution to continue at its current pace for more than a couple of years, advances in energy efficiency or software algorithms are necessary.

\subsection{Future Prospects}

\vaibify v1.0 stops at Level 3 because Level 4 represents an important phase change in the ladder's requirements: the data analysis must begin with raw observational data. While a substantial amount of important research has relied on published results without this trace, projects that reach Level 4 and higher are deemed to produce very high value results that require intense scrutiny and so should be traceable back to the raw data. Nonetheless, a path forward for \vaibify is clear and is presented in the rest of the subsection.

Levels 4 and 5 would be the next obvious features to add as they do not in principle require any community involvement. Many scientific publications include the raw data, so they can often be obtained without friction. New functionality could organize the prompts and supervision log that records the user's manual changes. Implementing Level 4 would also require combining these records with the git history and published artifacts, identifying gaps in the logs or other missing evidence, and a mechanism to verify the raw data. Similarly, \vaibify could be augmented to facilitate and monitor regeneration. Thus, should \vaibify be deemed useful, updates could enable researchers to reach Level 5 in the near future.

Level 6, however, is another crucial phase change as it would require community involvement and a system for verifying attestations. Online resources currently exist that are analogous to this requirement, and could be used as templates for this clearinghouse. This resource must be hosted and maintained, but such a hub could be invaluable for future scientists, both human and agentic, as they synthesize trustworthy results into a broader and deeper understanding of the universe. 

Once results regularly achieve Level 6, a community of agents could continuously update models and databases to reflect new observations and models, analogous to recent proposals to transform papers into agents \citep{Miao26}. This ecosystem could in principle drive scientific progress forward more quickly than humanity's collective intelligence, \ie the sum of human cognition would be slower than the sum of AI's cognition. Whether this state is achievable (or desirable) is not obvious at this time, but projects like \vaibify could ultimately help contribute to a system in which humans and AIs work together to advance science more rapidly than humans alone.

\end{document}